*Dedicated to the memory of Boris Klain*

# Toward a Phenomenological Theory of Earthquakes

A.V. Guglielmi

*Schmidt Institute of Physics of the Earth, Russian Academy of Sciences; Bol'shaya Gruzinskaya str., 10, bld. 1, Moscow, 123242 Russia; guglielmi@mail.ru*

**Abstract**

An attempt to construct a phenomenological theory of earthquakes is being undertaken by a small research team, which includes the author. We have only taken the first steps towards goal, but the results already obtained convincingly demonstrate the feasibility of the project. This paper is a kind of commentary on the path taken. The paper explains the essence and character of phenomenological theory. It also contains critical comments on alternative approaches to describing earthquakes. The theory is based on the concept of an earthquake source as a dynamic system, the state of which is described by a small number of phenomenological parameters. Within the framework of elementary theory, the source is described by one parameter, called the deactivation coefficient. The most important concept of the proper time of the source has been introduced, the course of which, generally speaking, differs from the course of world time. Dynamic equations describing the evolution of the source are given. The inverse problem of the source is posed, the essence of which is to calculate the phenomenological parameters based on earthquake observation data. Elementary phenomenological theory has already become a practical tool for the experimenter studying specific earthquake time series. The Appendix provides a list of previously unknown properties and patterns of earthquakes that we discovered within the framework of elementary theory. At the same time, it is noted that phenomenology itself does not lead us to a deep understanding of the essence of earthquakes. Sooner or later, the results of a complex of geophysical and physical-mathematical disciplines will lead to such an understanding. We place particular hopes on the theory of phase transitions and the mathematical theory of catastrophes.

## 1. Introduction

What properties should a satisfactory theory of earthquakes have? A small research team studied this issue and concluded that at this stage in the development of earthquake physics, an effective phenomenological theory is needed. The theory must be effective in the sense that it would open up new possibilities for processing and analyzing observations in order to search for previously unknown properties and patterns of earthquakes. In the papers [Gugllitlmim, 2015, 2017; Zavyalov et al., 2022; Guglielmi et al., 2023, 2025a,b; Zotov, Guglitlmi, 2025; Zotov et al., 2025], generalizing our experience, the path to a phenomenological theory of earthquakes is outlined. Only the first steps have been taken towards the goal. We managed to formulate an elementary phenomenological theory, which turned out to be quite effective (see Appendix $A$).

Methods of analyzing observation data developed on the basis of elementary theory have not yet received wide application in the practice of experimental earthquake research. This is partly due to the unusual character of the theory. This paper briefly outlines the elementary theory, comments on its unusual properties, and lists the results of studies of foreshocks, mainshocks, and aftershocks using methods developed on the basis of the theory.

## 2. Basic principles of elementary theory

The object of our study is the earthquake source, i.e. a stress-strain rock mass in which a main rupture spontaneously occurs, generating the main shock of the earthquake. Within the framework of elementary theory, the source is considered as a dynamic system. The state of the source is described by the phenomenological parameter $\sigma(t)$, which we will call the deactivation coefficient. The function $\sigma(t)$ is considered continuous and piecewise smooth.

We do not have a deductive theory of earthquakes within which the deactivation coefficient could be derived logically. Therefore, we should specify the procedure for calculating the values of $\sigma(t)$ from observation data of foreshocks, mainshocks and aftershocks.

The transition from a discrete sequence of underground impacts to continuous functions describing the evolution of the source is not simple and ambiguous. For example, when describing the relaxation of a source after the formation of a main rupture in it, Fusakichi Omori used the concept of instantaneous aftershock frequency $n(t)$ [Omori, 1894]. The values of the continuous function $n(t)$ are calculated by averaging over small intervals of universal time containing a sufficient number of aftershocks. We will call such a representation of the dynamics of the source the $t$-representation.

Our idea is that there is a proper time of the source, the course of which, generally speaking, differs from the course of world time. We explain the possible difference between proper time and world time by the non-stationarity of the conditions in which the dynamic system simulating the earthquake source is located.

The transition from world time to the proper time is ambiguous. We will consider two options: dimensionless time $s$ and time $\tau$ ,, which has the usual dimension. In the first case we will talk about the $s$-representation, and in the second case about the $\tau$-representation of the source dynamics.

Let us synchronize the proper time $s$ with the world time $t$ as follows. We will number the sequence of earthquakes with natural numbers $k = 1, 2, 3, ..$ and determine the world times $t_k$ for each event, for example, according to the USGS/NEIC catalog. Now let us imagine underground shocks as the ticking of an imaginary underground clock, the course of which is in some way coordinated with the course of dynamic processes in the source. The interval of instantaneous proper time between adjacent strikes is equal to 1, regardless of the value of the interval of world time $T_k = t_{k+1} - t_k$. The average interval of world time between two adjacent beats is equal to $T = \langle T_k \rangle$, where the angle brackets denote averaging over small intervals of proper time containing a number of discrete events sufficient to calculate $T$ with a given accuracy. Finally, we introduce the coordinate plane $(s, t)$, plot the points $t_k$ on it, and calculate a continuous piecewise smooth function $t(s)$ approximating these points. The synchronization function $t(s)$ can be used to calculate the mean universal time interval between adjacent earthquakes: $T = dt/ds$.

It should not be overlooked that the continuous proper time $s$ exists only in our imagination, giving us the opportunity to use the analysis of infinitesimal quantities to calculate the average interval $T$ between successive earthquakes. The only real, truly existing objects are points $t_k$ on the coordinate plane $(s, t)$. For them, by the way, there are finite arithmetic operations that give $T$, but we prefer to use the ideal representation of continuous proper time in order to use the apparatus of mathematical analysis.

We would like to briefly explain how the theory is formalized. Let us choose a formula that will serve as a stone for constructing the formal structure of elementary theory. We postulate the dependence of the deactivation coefficient on the proper time in the $s$-representation:

$$\sigma = \frac{d}{ds}\ln T. \qquad (1)$$

By our axiom we have already predetermined that the frequency of earthquakes $n = 1/T$ exponentially depends on the proper time of the source. Indeed, it is obvious that the dynamics of the source is described by a linear differential equation

$$\frac{dn}{ds} + \sigma n = 0. \qquad (2)$$

The solution has the form

$$n(s) = n_0 \exp\left(-\int_0^s \sigma(s')ds'\right). \qquad (3)$$

Here $n_0 = n(0)$ is the initial condition specified when formulating the Cauchy problem.

The synchronization function $t(s)$ is reversible, so it can be used to represent the deactivation coefficient as a function of world time

$$\sigma(t) = \frac{d}{dt}T(t), \qquad (4)$$

and introduce the proper time $\tau$ using formula

$$\tau(t) = \int_0^t \sigma(t)dt. \qquad (5)$$

In the $\tau$-representation, the evolution of the source is described by the simplest nonlinear differential equation

$$\frac{dn}{d\tau} + n^2 = 0. \qquad (6)$$

The solution of the evolution equation

$$n(\tau) = \frac{n_0}{1 + n_0 \tau} \tag{7}$$

indicates a hyperbolic dependence of the earthquake frequency on the proper time in the $\tau$-representation. It is curious to ask what meaningful meaning does the hyperbolicity of the continuous function (7) correspond to in the real world of discrete events? For simplicity, we set $\sigma = \text{const}$ and immediately discover that the intervals of world time between adjacent earthquakes form an arithmetic progression.

The structure of our theory is unusual. In this regard, the following comment may be helpful. It is quite clear that it would be more common to postulate not formula (1), but the evolution equation (2). Formula (3) would then be a solution to the direct problem, but it would be completely useless. The point is this. It is imperative to take into account that the phenomenological parameters of the theory must be calculated within the framework of a more fundamental theory, or, if this is not possible, as in our case, then a way to measure the parameters experimentally must be indicated.

The point is that our phenomenological theory is aimed at solving not the direct, but the inverse problem of the earthquake source. The inverse problem is to calculate the deactivation coefficient based on earthquake registration data. If we take the dynamic equation (2) as the basis of the theory, then formula (1) will not be a postulate, but a correct solution to the inverse problem.

So, if we were to briefly characterize the new understanding of the theory that our research initiated, we could perhaps say the following: The state of the source is described by the deactivation coefficient, which is calculated based on observation data of a discrete sequence of underground impacts by solving the inverse problem of evolution occurring over the proper time of the source.

## 3. Discussion

Our elementary theory, of which we have given an outline here, has laid a solid foundation for the study of earthquakes by new means of analysis. Having defined the source as a dynamic system and having identified the concepts of the deactivation coefficient and proper time, we have gained a clear understanding that not the direct, but the inverse problem should be the focus of attention when solving the phenomenological equations of the evolution of the source. The Appendix lists some results of an experimental study of earthquakes based on elementary theory. A detailed presentation of the results can be found in the review papers cited in the Introduction. We believe that work in this direction is promising. Particular attention should be paid to studying the phase trajectory of source in the extended phase space $(\sigma, \theta, s)$, where $\theta = d\sigma/ds$.

All attempts to go beyond the elementary theory have not yet brought success. Natural generalizations of the equations of evolution within the elementary theory were made by adding new terms to them. In an effort not to violate the basic tenets of the elementary theory, we made minimal additions motivated by various geophysical considerations. For example, equation (6) implies that the aftershock frequency asymptotically approaches zero with the passage of universal time. This conclusion is not supported by observations, which indicate that after relaxation, the source usually switches to background seismicity mode. By adding a linear term $\gamma n$ to the right-hand side of the quadratic equation (6), we arrive at the Verhulst logistic equation. Let's write it in $t$-representation

$$\frac{dn}{dt} = n(\gamma - \sigma n). \qquad (8)$$

When $t \to \infty$, the solutions of equation (8) give the background seismicity frequency $n_\infty = \gamma/\sigma$. In principle, from here it is possible to calculate a new phenomenological parameter $\gamma$ based on data on the frequency of earthquakes in the background seismicity regime.

From the ordinary differential equation of evolution one can go to a partial differential equation by making the substitution $n(t) \to n(\mathbf{x}, t)$ and adding the diffusion term $D\nabla^w n$ to the logistic equation. Thus, we obtain the Kolmogorov–Petrovskii–Piskunov equation, which describes the propagation of nonlinear diffusion waves. The concept of diffusion waves allows us to understand in general terms the surprising phenomenon of aftershock divergence discovered in [Zotov, Zavyalov, Klain, 2020].

Mentally operating with a continuous wave through a finite discrete sequence of underground impacts that exists in reality brings us to the limit of the possibilities of elementary theory. The introduction of a third phenomenological parameter $D$ in itself requires additional measurements, for example, measuring the propagation velocity of a diffusion wave. Here the procedure of transition from discrete to continuous can become more reliable if we manage to include Umov's concept of the rate of energy flow in a solid body into the phenomenological theory. Efforts in this direction are justified, since we have no doubt whatsoever about the existence of directed energy flows at the earthquake's source. Moreover, there is reason to assume the activation of the vortex movement of energy flows in the source before the main shock of the earthquake.

Let us return, however, to the evolution equation of the simplest form (6) and add a free term to the right-hand side. If the free term is chosen as a delta-correlated random function with zero mean, we obtain the stochastic Langevin equation, suitable, in particular, for studying fluctuation phenomena at the source.

In conclusion of the discussion, we note that the elementary theory eliminates a shortcoming that existed in the physics of earthquakes. The Hirano-Utsu law is widely used in aftershock analysis, according to which the frequency of aftershocks decreases according to a power law with the passage of world time (see, for example, [Hirano, 1924; Jeffreys, 1938:; Utsu, 1957; Utsu, Ogata, Matsu'ura, 1995; Ogata, Zhuang, 2006: Rodrigo, 2021; Salinas-Martínez, at al. 2023]). According to tradition, established by long-term practice, the exponent is denoted by the symbol $p$. The value of $p$ varies from case to case within limits, approximately from 0.5 to 1.7. The Hirano-Utsu law is an infinite set of statements, since we can substitute many values of $p$ into it, for example, $p = 0.8$, or $p = 1.1$. However, experience in studying aftershocks using phenomenological theory methods has shown that the power law is satisfied only when $p = 1$ in one of the three relaxation phases of the source. At $p = 1$, the Hirano-Utsu law coincides with the Omori law. For $p < 1$ ($p > 1$), the deactivation coefficient decreases monotonically (increases monotonically) over time if the aftershock frequency is approximated by a power law. However, many years of experience in studying aftershocks in light of elementary earthquake theory confidently demonstrates that the deactivation coefficient of the source is not a monotonic function of time.

Thus, it appears that the Hirano-Utsu law, unlike the Omori law, is not realized during the relaxation process of the source after the main shock. Omori's law is observed, and observed strictly. But it is not holistic. Omori's law is satisfied over a limited time interval during relaxation of the source. We call this time interval the *Omori epoch*. It is worth mentioning here that the main step forward in the formation of an elementary theory was the discovery of the fact that the Omori epoch exists in each of the aftershock series we analyzed. As a result, the elementary theory gained credibility in its field and became a practical tool for the experimenter studying the patterns of specific earthquake time series.

It is quite obvious, and it goes without saying, that phenomenological theory by itself does not lead us to a deep understanding of the essence of earthquakes. Sooner or later, the results of a complex of geophysical and physical-mathematical disciplines will lead to such an understanding. We place particular hopes on plate tectonics and the mathematical theory of catastrophes.

## 4. Conclusion

In conclusion, it would be worthwhile to indicate the general direction of development of the phenomenological theory of earthquakes. But our research group, which includes Boris Klain, Alexey Zavyalov, Oleg Zotov, and the author of this paper, is actively searching for ways to develop the theory. The possibilities of phenomenology are extremely interesting and diverse [Zotov et al., 2025]. We haven't made any key decisions yet. So, instead of describing the future, I'll briefly tell the story of how, already in my old age, I first became interested in earthquake physics.

Together with O. Zotov, we drew attention to the cumulative impact of converging seismic waves on the earthquake source. In the course of studying the cumulative effect, we discovered modulation of global seismicity by spheroidal oscillations of the Earth. We were not experts in earthquake physics. Therefore, we turned to the renowned seismologist A. Zavyalov, and together with him, we have convincingly substantiated the existence of both effects through a thorough analysis of the aftershocks of the Sumatra-Andaman earthquake and the Tohoku earthquake. All work was carried out in close creative collaboration with B. Klain, who possessed deep knowledge of mathematics. This is how a creative team was formed that studies earthquakes using phenomenological methods.

*Acknowledgments*. I would like to express my gratitude to B. Klain, A. Zavyalov, and O. Zotov for their joint work. Only through concerted, common efforts were we able to overcome significant difficulties and dramatic collisions, and bring our new method of processing and analyzing earthquakes to a state in which it began to yield tangible results in the search for and discovery of new properties and patterns.

**Appendix *A***

*List of research results within the framework of elementary phenomenological theory*

**1.** A classification of main shocks within tectonic earthquake triads has been proposed. Each main shock belongs to one of three classes, with each class divided into two species. Four previously unknown main shock species have been discovered.

**2.** For foreshocks of mirror triads, Zotov's law has been established, similar to Bath's law, which was established earlier for aftershocks of classical triads.

**3.** The phenomenon of foreshock convergence and aftershock divergence has been detected. The maximum foreshock convergence velocity (85 m/h) of the catastrophic Tohoku earthquake ($M = 9.1$) has been measured.

**4.** Rotational movement of foreshock epicenters around the main shock epicenter was detected

**5.** Three phases of earthquake source relaxation have been identified. The source deactivation coefficient is zero in the initial phase, assumes a constant positive value in the main phase, and undergoes chaotic variations over time in the recovery phase.

**6.** A regular decrease in the deactivation coefficient in the main relaxation phase of the source with an increase in the magnitude of the main shock was established.

**7.** The cumulative effect of the round-the-world seismic echo has been predicted and discovered.

**8.** Modulation of global seismicity by spheroidal and toroidal oscillations of the Earth has been discovered.

**9.** Anthropogenic "weekend" effect in global seismicity has been discovered.

**10.** It has been established that the Hirano-Utsu law is not applicable to the description of aftershocks.

## Appendix *B*

*Pilot analysis of foreshocks in the s-representation of the Tohoku earthquake source dynamics. A detailed analysis will be presented in the paper [Zotov et al., 2025]*

The main shock of the Tohoku earthquake with a magnitude of *M* = 9.1 occurred on March 11, 2011. In the 200 days before the main shock, 166 foreshocks were recorded.

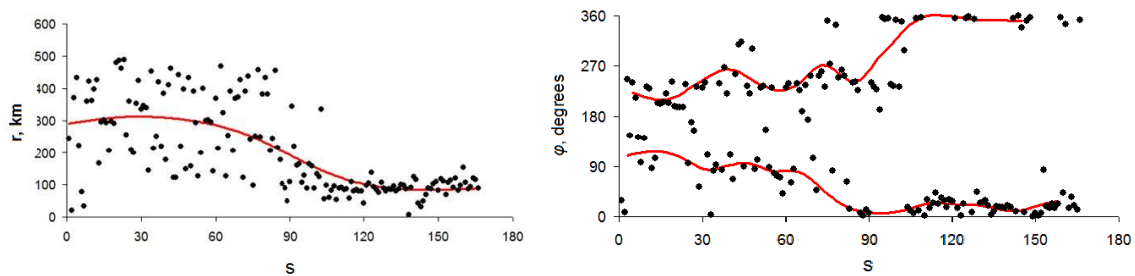

Radial (left) and azimuthal (right) dependence of foreshock epicenters on the proper time of the source. The center of the circular coordinate system is aligned with the mainshock epicenter. Azimuth is measured counterclockwise from the east. Red lines represent spline approximations of the experimental points.

The figure shows the distances and azimuths of foreshocks depending on the proper time of the earthquake source. We clearly see a tendency for foreshock epicenters to converge toward the mainshock epicenter.

The right panel demonstrates the spatial division of foreshocks into two populations. Indeed, the dots clearly line up along the two branches. Movement along the lower (upper) branch occurs clockwise (counterclockwise). Taking into account the convergence of foreshocks, azimuthal movements indirectly indicate the existence of two Umov energy flows in the source, converging towards the hypocenter of the future main shock and rotating in mutually opposite directions. Using the synchronization function *t*(*s*), we can estimate the radial transfer velocity, which amounts to several tens of meters per hour. The rate of rotational movement can also be estimated as follows: 0.7 degrees per day along the lower branch and 1.3 degrees per day along the upper branch.

Let us introduce the coordinate plane $(x, y)$ so that it touches the earth's surface, and align the origin of coordinates with the epicenter of the main shock. Let us determine the coordinates $(x_k, y_k)$ of the foreshock sequence, $k = 1, 2, .. 166$. Let $\Delta x_k = x_{k+1} - x_k$, $\Delta y_k = y_{k+1} - y_k$. Then the distance between adjacent foreshocks is $\Delta l_k = \sqrt{(\Delta x_k)^2 + (\Delta y_k)^2}$. The length of the broken path is

$$l_k = \sum_{i=1}^{k} \Delta l_i.$$

Here $k = 1, 2, .. 165$. Let us plot points $l_k$ on the coordinate plane $(l, s)$ and find a differentiable function $l(s)$ that approximates these points. Using function $l(s)$ and taking into account the synchronization function $t(s)$, we can calculate the foreshock velocity along the smoothed trajectory:

$$V = \frac{1}{T}\frac{dl}{ds}.$$